\documentclass[12pt]{article}

\usepackage{natbib}
\usepackage{hyperref}
\hypersetup{
  colorlinks=true,
  linkcolor=blue,
  filecolor=magenta,
  urlcolor=cyan,
  citecolor=blue,
}

\usepackage{ifpdf}
\usepackage{lscape}
\usepackage{amsmath}
\usepackage{amssymb}
\usepackage[margin = 1.2in]{geometry}
\usepackage{graphicx}
\usepackage{natbib}
\usepackage{subfigure}
\usepackage[para]{threeparttable}
\usepackage{booktabs}
\usepackage{multirow}
\usepackage{rotating}
\usepackage{setspace}
\usepackage{floatrow}
\usepackage{chngcntr}
\usepackage{epstopdf}
\usepackage{enumerate}
\usepackage{float}
\usepackage{placeins} 
\usepackage{CJK}
\usepackage{type1cm}
\usepackage{times}
\usepackage[marginal]{footmisc}
\renewcommand{\thefootnote}{}
\usepackage{pdflscape}
\usepackage{url}
\usepackage{color}
\usepackage[table,xcdraw]{xcolor}
\usepackage[misc]{ifsym}
\usepackage{authblk}
\usepackage{url}
\usepackage{hyperref}
\usepackage{multirow}
\hypersetup{
    colorlinks=true,
    linkcolor=red,
    filecolor=magenta,      
    urlcolor=gray,
    citecolor=blue
}
\usepackage{adjustbox}

\newcommand{\qlet}{\hspace*{\fill} \raisebox{-1pt}{\includegraphics[scale=0.06]{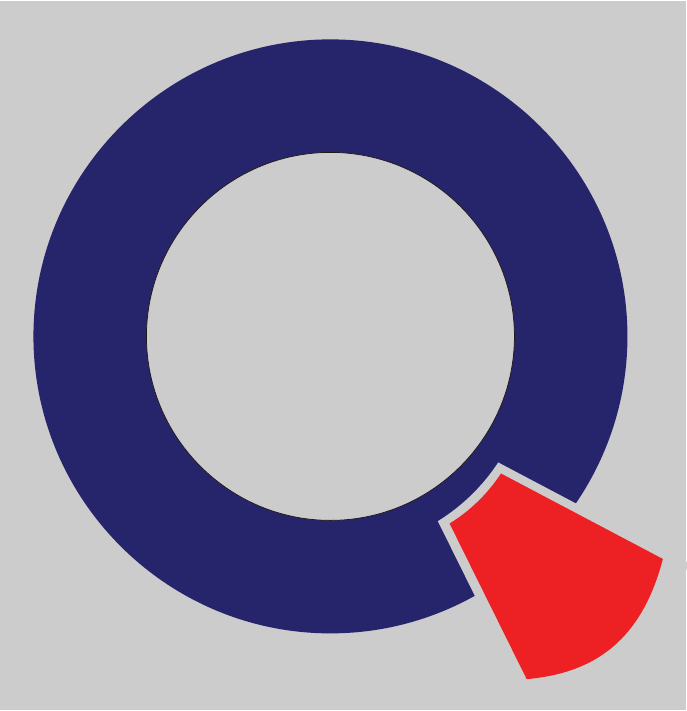}}}

\newtheorem{RProblem}{\hskip 2.5em Research Question}

\newenvironment{proof}[1][Proof]{\begin{trivlist}
  \item[\hskip \labelsep {\bfseries #1}]}{\end{trivlist}}

\newcommand{\qed}{\nobreak \ifvmode \relax \else
  \ifdim\lastskip<1.5em \hskip-\lastskip
  \hskip1.5em plus0em minus0.5em \fi \nobreak
  \vrule height0.75em width0.5em depth0.25em\fi}

\definecolor{navyblue}{rgb}{0.0, 0.0, 0.5}
\begin{document}

\title{A Machine Learning Based Regulatory Risk Index for Cryptocurrencies}
%

\author[a]{Xinwen Ni \thanks{Financial support of the European Union’s Horizon 2020 research and innovation program “FIN- TECH: A Financial supervision and Technology compliance training programme” under the grant agreement No 825215 (Topic: ICT-35-2018, Type of action: CSA), the European Cooperation in Science \& Technology COST Action grant CA19130 - Fintech and Artificial Intelligence in Finance - Towards a transparent financial industry, the Deutsche Forschungsgemeinschaft’s IRTG 1792 grant, the Yushan Scholar Program of Taiwan and the Czech Science Foundation’s grant no. 19-28231X / CAS: XDA 23020303 are greatly acknowledged. }}
\author[a,b,c,d,f]{Wolfgang Karl H{\"a}rdle}
\author[g]{Taojun Xie}

\affil[a]{\small{School of Business and Economics, Humboldt-Universit{\"a}t zu Berlin, Berlin, Germany}}
\affil[b]{\small{Sim Kee Boon Institute for Financial Economics, Singapore Management University, Singapore}}
\affil[c]{\small{W.I.S.E. - Wang Yanan Institute for Studies in Economics, Xiamen University, Fujian, China}}
\affil[d]{\small{Department of Probability and Mathematical Statistics, Faculty of Mathematics and Physics, Charles University, Prague, Czech Republic}}
\affil[f]{\small{Department of Information Management and Finance, National JiaoTong University, Taiwan}}
\affil[g]{\small{Asia Competitiveness Institute, Lee Kuan Yew School of Public Policy

National University of Singapore, Singapore}}

\maketitle
\pagenumbering{gobble}

\begin{abstract}
Cryptocurrencies' values often respond aggressively to major policy changes, but none of the existing indices informs on the market risks associated with regulatory changes. In this paper, we quantify the risks originating from new regulations on FinTech and cryptocurrencies (CCs), and analyse their impact on market dynamics. Specifically, a \textcolor{blue}{\bf C}ryptocurrency \textcolor{blue}{\bf R}egulatory \textcolor{blue}{\bf R}isk \textcolor{blue}{\bf I}nde\textcolor{blue}{\bf X} (CRRIX) is constructed based on policy-related news coverage frequency. The unlabeled news data are collected from the top online CC news platforms and further classified using 
a Latent Dirichlet Allocation model and Hellinger distance.
Our results show that the machine-learning-based CRRIX successfully captures major policy-changing moments. The movements for both the VCRIX, a market volatility index, and the CRRIX are synchronous, meaning that the CRRIX could be helpful for all participants in the cryptocurrency market. The algorithms and Python code are available for research purposes on www.quantlet.de. 

\vspace{0.2in} \noindent
\end{abstract}



\noindent

\begin{spacing}{1.5}

  \newpage{}
  \clearpage{}
  \pagenumbering{arabic}
  \setcounter{page}{1}
  
\FloatBarrier
  \section{Introduction}
\renewcommand{\thefootnote}{\arabic{footnote}}

Today, there are nearly 2,500 cryptocurrencies worth more than \$252.5 trillion trading in the market (\citealp{nu_coins}).
The original boom of cryptocurrencies occurred in an unregulated environment. Even as news outlets and investors paid closer attention to the market, regulators and international actors remained largely distant from the action, and prices continued to soar unabated. However, the situation changed since 2014. 

Regulations are designed to protect the
investors, to put a stop on money laundering, or to prevent the fiat currency from being
crowded out. Despite these good wills, speculation and implementation of regulations have resulted in volatile price movements in the cryptocurrency markets. Recent incidents,
including China's ban on cryptocurrency exchanges and the rumours of Korea doing the
same, have caused major sell-offs and losses among investors. It is therefore important to
identify the extent to which new regulations and speculations on them have affected the
cryptocurrency markets. Ignoring this source of risk, regulators could end up with self-destroying
outcomes and create thus systemic risk bias.

In this paper, we aim to quantify the risks originating from introducing regulations on the Cryptocurrency (CC) markets and identify their impact on the cryptocurrency investments.  In order to measure the regulatory risk, particularly the effects of regulations, some researchers considered  event-study methods (\citealp{binder1985measuring}; \citealp{buckland2001political};  \citealp{binder1985measuring}; \citealp{binder1985measuring}; \citealp{schwert1981measuring} and etc.). However, cryptocurrency market is young and so different from other financial market that the previous regulatory event may not appear again, such as a certain country ban the market. Therefore, we need a measurement tool, which is able to represent the risk level, be comparative and track the changes over time. An index matches all those requirement. 

 Indices have been applied to track the Cryptocurrency markets already. The CC index developed in \cite{trimborn2018crix}, known as CRIX \footnote{seen in \url{crix.berlin}, or \url{thecrix.de}} is a benchmark and tracks the price movements in the
CC markets on a daily basis. A volatility measure, VCRIX (\citealp{kim2019vcrix}), similar to VIX, is also presented.
there to reflect the market's volatility 
However, even though the VCRIX shows jumps that are to some extent attributable
 to political decisions, neither of these indices and
other indices, like e.g. CCI30 \footnote{The CCi30 is a rules-based index designed to objectively measure the overall growth, daily and long-term movement of the blockchain sector. It does so by tracking the 30 largest cryptocurrencies by market capitalization, seen in \url{http://cci30.com}} directly address regulatory risks although it
plays an important role for the future of CCs and in addition might be created by text
mining techniques from a sufficiently rich corpus. 


The here proposed Cryptocurrency Regulatory Risk Index (CRRIX) will be constructed by evaluating regulation-related news articles. The indices introduced in \cite{baker2016measuring} reveal economic policy uncertainty. Similar to their indices, our index is also based on the policy-related news coverage frequency. Unlike their algorithm, which involved a meticulous manual process to label a pool of 12,000 articles, we pursue a Machine Learning (ML) technique to classify policy-related news in our data. We use the support vector machine (SVM), a widely used text classifier, as benchmark and apply Latent Dirichlet Allocation (LDA) method to capture the distances between articles and further classify the unannotated news according to their similarity to policy-related news.  

%
%

The Regulatory Risk Index (CC) will subsequently be used to analyse the association between regulatory risk and
market activities. The algorithms have been programmed in Python. All numerical calculations are available for research purposes on \url{www.quantlet.de} and also
on Github in the organization QuantLet , see \cite{borke2018q3}. 

In the next Section \ref{sec:researchQ}, we discuss the background and research questions. The Section \ref{sec:data}
presents the data in detail and shows the basic statistics. The Section \ref{sec:method} enters into the methodology and the Section \ref{sec:Results } presents results from  LDA model and the time series of our regulatory risk index. Finally, 
Section \ref{sec:conclusion} concludes.



  \section{Background and Research Questions}
  \label{sec:researchQ}
  In this section, we first generally review the development of cryptocurrency. Based on that, we display the picture of trends of regulatory dynamics in the CC market. The research questions touch the identification of the regulatory risk, the construction of the regulatory risk index, and the interaction of the regulatory risk index with market variables.

  \subsection{Development of Cryptocurrency}
  The history of digital or programmable monies can be traced back to as far as four decades ago, although the word ``cryptocurrency'' became popular only recently. In the 1980s, the concept of E-cash was introduced by David Chaum in a paper entitled ``Blind Signatures for Untraceable Payments" (\citealp{fiorillo2018bitcoin}). Subsequently, DigiCash, B-Money, and Bit Gold were proposed by Chaum, Wei Dai, and Nick Szabo in the 1990s, respectively. Most of these digital monies did not survive as they failed to address the practical issues of ``double spending'' and ``third-party trust''.

The milestone development in this field came in after the global financial crisis. \citeauthor{nakamoto2019bitcoin}, in the seminal white paper, \citeyear{nakamoto2008bitcoin}, proposed the bitcoin. This is a peer-to-peer electronic cash system implemented via the blockchain technology, with the participation of a network of computer owners known as miners. The blockchain technology, later known as the distributed ledger technology, or DLT, ensures that transaction records are easy to be updated but costly to be changed, avoiding the ``double spending'' issue. Miners, after solving complex mathematical puzzles, are rewarded with a predetermined amount of bitcoins. The amount of the reward can only be amended with the agreement from majority of the miners. Such a mechanism avoids the ``thrid-party trust'' problem in fiat monies, whose issuance depends on the central banks' sole discretion. Since its birth, the BTC model has defined the meaning of ``cryptocurrency'', which now typically refers to a decentralized digital network that facilitates secured transactions using cryptographic methods.

\begin{figure}[!h]
    \centering
    \includegraphics[height = 0.9\textheight]{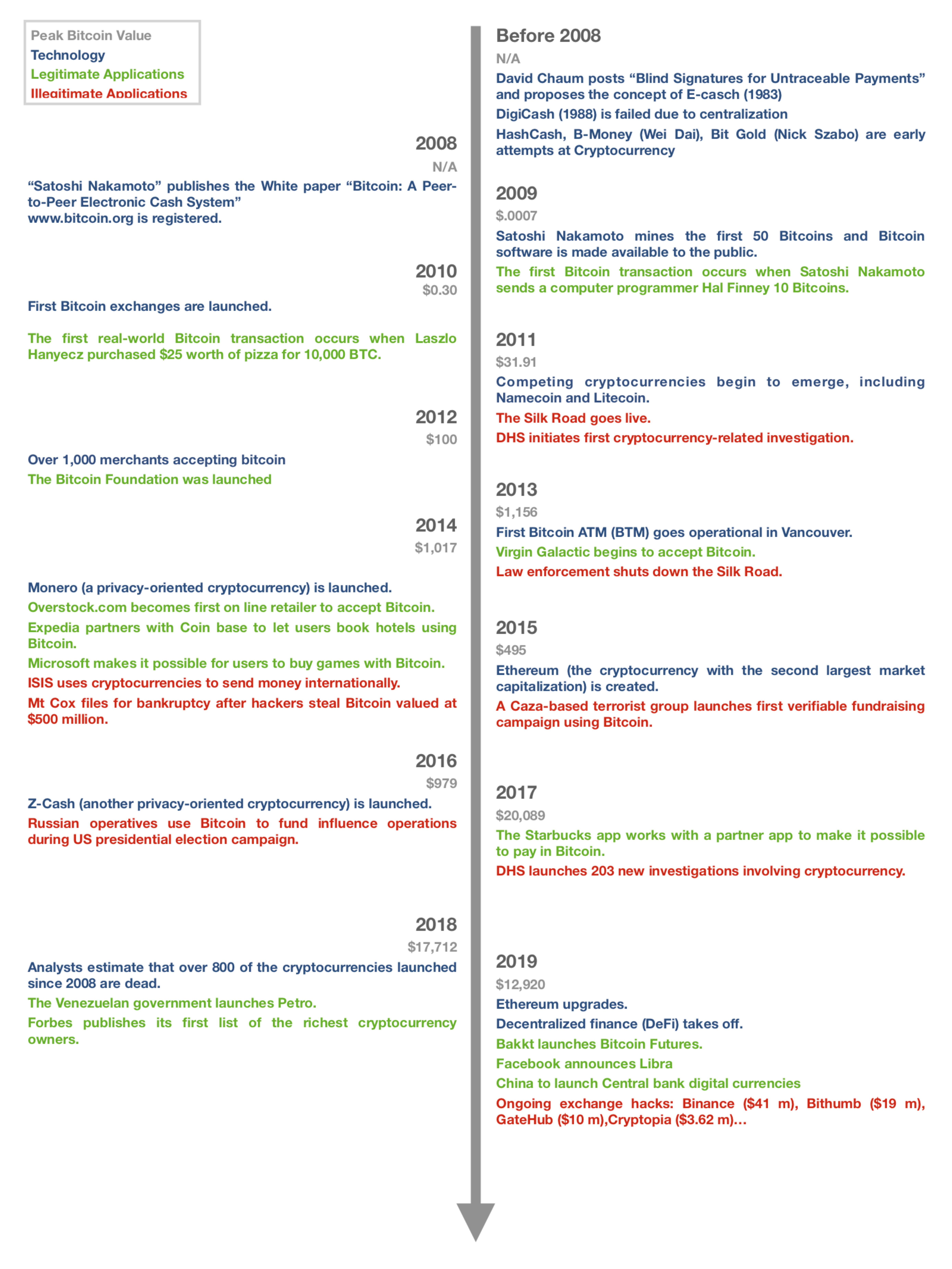}
     \caption[Cryptocurrency timeline]{Cryptocurrency timeline} 
    \label{fig:timeline_new}
\end{figure}

 
 Bitcoin has sparked a series of events since 2009. In Figure \ref{fig:timeline_new}, we show a timeline of a few major events in the last decade. Some events pertained to innovations that aimed at improving the technology behind cryptocurrencies (highlighted in blue). For example, competing CCs, also known as altcoins, began to emerge in 2011. The events highlighted in yellow are the ones involving CCs being used in legitimate real-life applications. A well-known example was the two pizzas in 2010 that cost 10,000 bitcoins. Through this series of events, the general public became aware of the strengths and weaknesses of CCs. As good and bad news took turns to be reported, the price of bitcoin, highlighted in gray, and that of the other CCs also experienced volatile movements.  Notably, events relating to losses of cryptocurrency exchanges have been associated with the largest price movements. For instance, as Mt.~Gox went bankruptcy in 2014 after losing over 850,000 bitcoins, the price of bitcoin fell from over \$1,000 to the around \$400. This event has been foreseeable through textual analysis of BTC blog and other solid media channels. \cite{linton2017dynamic} and the chapter 3 of \cite{hardle2017applied} employ a technique similar to ours to evaluate discussions in social media. A recent price movement was an upswing of 1300\% in year 2017, followed by a fall of more than half in May 2018. 
 
 \subsection{Regulations of the Cryptocurrency Market}
 Among the features of cryptocurrencies, anonymity has been the most controversial one. Users of cryptocurrencies like this feature for it is difficult to trace one's spending history, but regulators dislike it for the exact same reason. We thus see an interesting interaction here. Users have proposed numerous improvements to enhance anonymity. Zcash and Monero were designed to facilitate anonymous transactions. At the same time, incidents such as the Silk Road going live and terrorists using cryptocurrencies for remittance got the regulators on the toes. Regulators, on the one hand, insisted on know-your-customer (KYC) measures to trace any illegitimate transactions, but on the other hand, prepared to launch their own cryptocurrencies \citep{barrdear_macroeconomics_2016,george_central_2020}. Fighting against illegitimate transactions became one of the first tasks for the cryptocurrency regulators.
 
 
Trading activities at the exchanges was the next issue that regulators reacted to. The anonymity feature and a lack of regulation at the cryptocurrency exchanges cultivated illegal and unethical behaviors, such as money laundering, pump-and-dump activities and scams. Ignorant users of the cryptocurrency exchanges faced high risks while trading. Responding to these, starting in 2011, the US Treasury Department's Financial Crimes Enforcement Network (FinCEN) began oversight of cryptocurrency exchanges, transmitters, and administrators under the Bank Secrecy Act related to anti-money laundering and combating the financing of terrorism (AML/CFT) (\citealp{lee2018handbook}). In the same year, the United States Department of Homeland Security (DHS) initiated first investigation relating to cryptocurrency. The number of cases raised to over 200 in 2017 (seen in Figure \ref{fig:timeline_new}).

\begin{figure}[!h]
    \centering
    \includegraphics[width = 12cm]{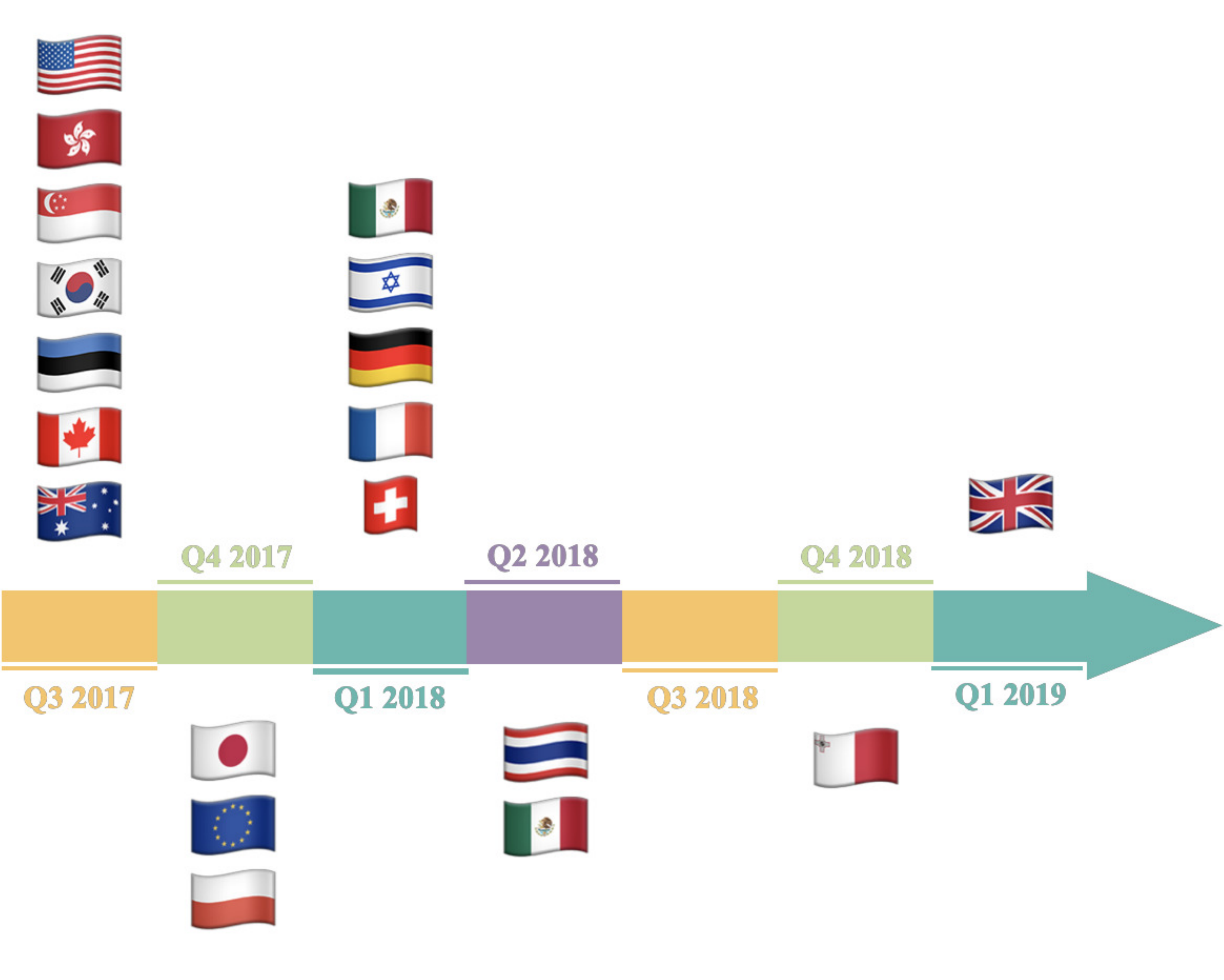}
     \caption[timeline_guidance]{Timeline for Cryptocurrency Guidance} 
    \label{fig:timeline_guidance}
\end{figure}

However, there has been a dilemma in regulating the cryptocurrency space. While it was important to protect retail investors and to prevent unlawful transactions, new technology driving the development of cryptocurrencies needed to be incubated until the ecosystem was matured. It was then imperative for the regulators to move towards systematic governance. In Figure \ref{fig:timeline_guidance}, we show a timeline of countries' publications of guidance on the cryptocurrency space. 


%
%
In order to apply the existing law and regulations, the very first step for the governors is to identify the cryptocurrency nature as means of holding, transferring or investing ``money". Since the birth of cryptocurrency, the debate, whether BTC, or generally cryptocurrencies, are currency or asset, eolved with the growth of the market (\citealp{glaser2014bitcoin}, \citealp{baur2018bitcoin}). Despite using ``currency'' as the name name, most countries that have permitted cryptocurrencies view these monies as assets. In 2014, the US Internal Revenue Service (IRS) stipulated that virtual currency is considered property for the purpose of federal tax''. Purchases using cryptocurrencies as the media of exchange are considered barter trades, or exchanges between properties and services (\citealp{blandin2019global}). The Swiss Financial Market Supervisory Authority (FINMA), in 2018, published guidelines for initial coin offerings (ICOs). In those guidelines, tokens were categorised into three types based on their economic functions: payment tokens, utility tokens and asset tokens (\citealp{caytas2018regulation}). 

Regulatory responses to cryptocurrencies vary from strict bans to government adoption. Most of these regulations were designed to protect retail investors or to ensure that cryptocurrencies are not used for illegal activities.
 The Financial Action Task Force (FATF), recommended that regulations be implemented to prevent the use of cryptocurrencies in money laundering and terrorist finance (\citealp{gold2019cryptocurrency}). At the G20 Summit in 2018, the FATF urged on all countries to take necessary preventive measures towards the misuse of cryptocurrencies. In the European Union, by 2020, all member states will implement AML/CFT\footnote{AML/CFT: Anti-money laundering / combating the financing of terrorism.} rules to cryptocurrency exchanges and wallet operators (\citealp{houben2018cryptocurrencies}). On the contrary, some countries began to change their attitude towards the adoption of DLT. In China, the use of BTC as currency for financial institutions was prohibited in 2013 (\citealp{glaser2014bitcoin}), but in Octobor 2019, the president of China announced that the country will encourage enterprises to seize the opportunity in the up-and-coming technology. Subsequently, the People's Bank of China announced that it would launch digital Yuan, commonly known as the central bank digital currency (\citealp{chinanewc_currency}). Such support for DLT lead to a new round of debate globally.

This disparity in regulatory approaches creates interesting dynamics in the cryptocurrency markets. As poited out earlier, good and bad news is likely to induce different movements in the price of cryptocurrencies. As more regulations are on their way, and because the cryptocurrency market is globally unified, policy changes in one country or even the rumor about the attitude adjustment of one government, would be widely discussed in the news platform and bring turbulence to the whole market. Under these circumstances, and in consistent with \cite{Auer_Cryptocurrency_2020}, the construction of a regulatory risk measurement for the whole market is necessary and the news data, which includes information from different sources and countries, would be helpful.

\subsection{Research Questions}

For the most cases, the regulation changes do not come out from nowhere. A good example is the Fed's rate cut (or hike). Before the Fed's announcement of changes, there are discussions and predictions about the Fed's move in newspapers. Numerous studies have attempted to examine whether the text mining technology would contribute to the forecasting for financial markets (\citealp{nassirtoussi2014text}), e.g. \cite{ghiassi2013twitter}, \cite{geva2014empirical}, \cite{tu2018information}. But not many use textual data to analyze financial regulatory risk. \cite{gulen2016policy}, \cite{baker2016measuring} and \cite{kang2013oil} argue that news from newspapers could be a good indicator for macro policy uncertainty. 
 
As discussed before, indices have been introduced to trace the movement of CC market, but none of them addresses regulatory risks which plays an important role for the future of CCs. We try to, in this paper, quantify the risks brought by introducing policies on the CC market and further discuss its impact on the CC investment.  A regulatory risk index for the CC market, which could serve as a tool for passive investors, for the fund manager, and even for policy-makers.

There are mainly three kinds of indices as to the data sources which were applied to construct the index. 
First, and most commonly, some indices use real data, e.g. VIX, S\&P 500, and DAX, which employed real market price or volume data. Second, some are based on a regular survey, e.g. IFO business Climate Index and Purchasing Managers' Index (PMI), which is relied on a monthly survey of supply chain managers. Recently, the third source, news data, or generally speaking the text data, becomes popular, e.g. Thomson Reuters MarketPsych Indices \footnote{seen in \url{https://www.refinitiv.com/en/products/world-news-data/}}, sentiment Indices, which are standard input to trading desks and are provided by a variety of news data channels. The majority of these indices are based on text mining tools that establish a dictionary and then calculate in
a bag of words approach the frequency of key words with identifiable sentiment, like good,
bad etc. Such input for asset trading and risk management is a refined data source that in
many cases contains useful directional information. \cite{baker2016measuring} introduced an index to reveal economic policy uncertainty based on newspaper coverage frequency. They calculated the number of articles which contained ``economic" or "economy"; ``uncertain" or ``uncertainty"; and one or more of ``congress", ``deficit", ``Federal Reserve", ``legislation", ``regulation" or ``White House" from 10 large newspapers. Their index successfully represented movements in policy-related economic uncertainty and they also found that policy uncertainty rise stock price volatility.
  
Unlike \citeauthor{baker2016measuring}'s algorithm, which involved a meticulous manual process to label a pool of 12,000 articles, we propose a machine learning based approach to classify policy-related news in our data.  
With a sufficiently rich corpus, it might be able to conduct the regulatory risk index by some NLP techniques, such as topic modelling methods. We are also curious about whether the policy-related uncertainty index could be helpful for the market participants. Therefore, the research questions discussed in this paper are as follows:

\begin{RProblem} 
How to indentify the regulatory risk for Cryptocurrencies? 
\end{RProblem}

\begin{RProblem} 
 How to construct an index of regulatory risk for Cryptocurrency market based on news data?
\end{RProblem}   

\begin{RProblem} 
 What is the impact of regulatory risk to the market?
\end{RProblem} 

%
  \section{Data}  
  \label{sec:data}
\subsection{News from Top Cryptocurrency Online Platforms}

Since CCs are frequently mentioned in newspapers only for the very recent years, we choose to use news data from the top online cryptocurrency news platform (\citealp{guides2018top}). The representative news platform Coindesk \footnote{\url{https://www.coindesk.com}} and Bitcoin Magazine \footnote{\url{https://bitcoinmagazine.com}} were considered in this paper, because  both of them are not only pioneers and leaders in the market but also they offer news data which could trace back to the beginning of BTC's boost. 

Coindesk is a news website with a particular focus on Blockchain, Bitcoin and Cryptocurrencies
as a whole. The site launched in April 2013 and released close to 25000 articles. The articles have been classified already in categories: markets, technology, business, policy \& regulation and people. The aim of this paper is to introduce a policy uncertainty index by calculating the frequency of policy-related news. The pre-classified news data from Coindesk perfectly matches our demand and will be further applied as training data for the ML models.  The textual data from the source was collected
via a dynamic web scraper. 

After checking for duplicates, eventually we keep 16,528 articles from 01 April 2013 to 18 July 2019. The data covers 76 months, 329 weeks and 2300 days. Out of the total over 16,000 articles, 2,468 are marked as policy-related news. The data is available for further research at the Blockchain Research Center (BRC) \footnote{\url{https://hu.berlin/BRC}} and on \url{Quantlet.de}. 
 
\begin{figure}[!h]
    \centering
    \includegraphics[width = 12cm]{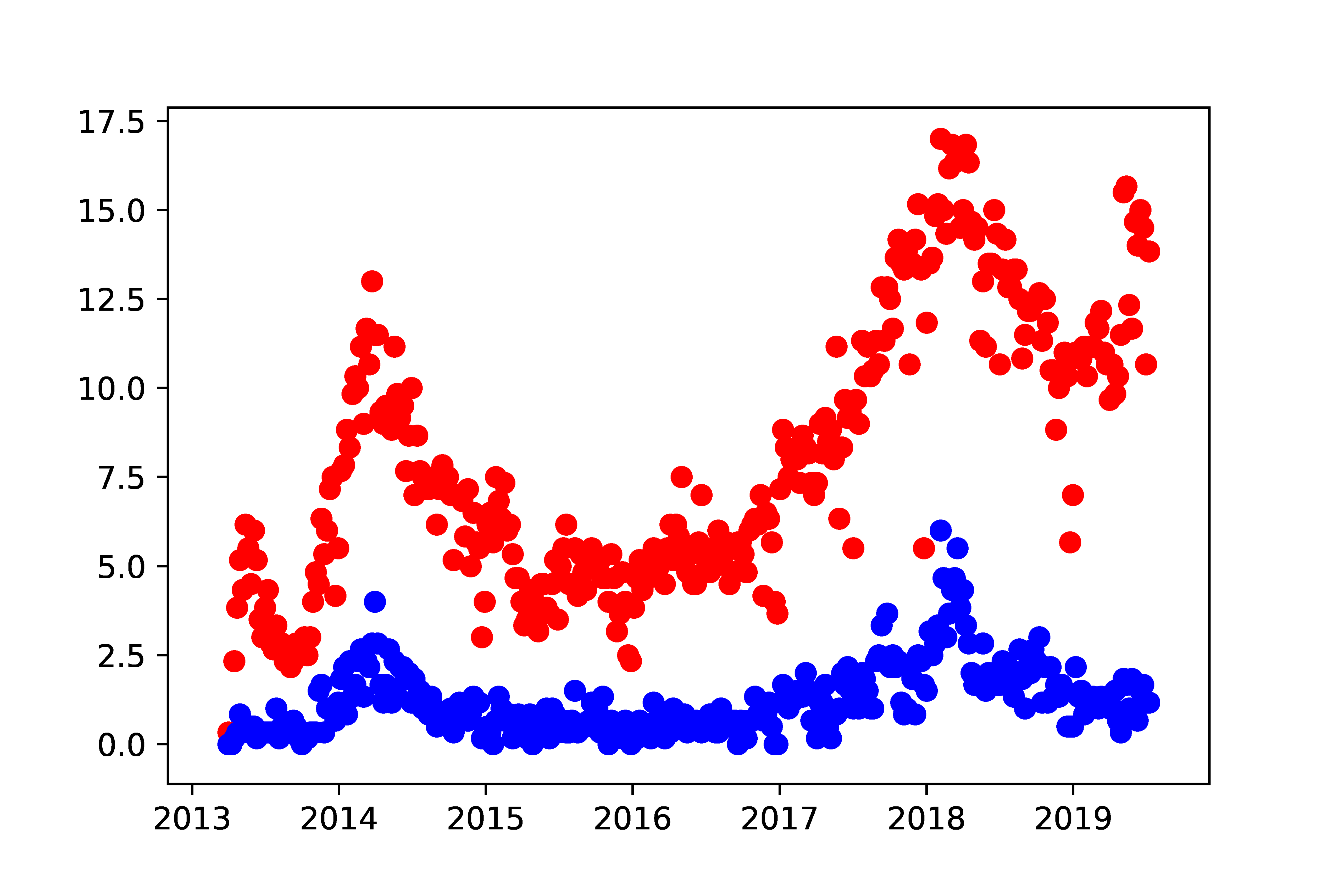}
     \caption[num_news_regs]{Weekly Average Number of total News (in \textcolor{red}{red}) and of Policy-Related News (in \textcolor{blue}{blue})} 
    \label{fig:num_news_regs}
\end{figure}

Figure \ref{fig:num_news_regs} represents the average number of news per week related to policy and the average number of all news. The number of daily articles increased in 2014 and 2018 both in total and in regulation-related term. In those years, the price of Bitcoin underwent volatile movements, declining
by more than 70 \% in 2014 in 2018. Before these massive corrections, the end of 2013 and 2017
marked periods of price discovery and all-time highs in USD valuation being broken every
other day. During the same periods, number of blockchain-related news articles increased, indicating 
growing interests in blockchain and distributed ledger technology. There is no doubt that, simultaneously, the market attracted policy-makers' strong attention as well.

\href{https://github.com/QuantLet/Regulatory-risk-cryptocurrency}{\qlet{CRRIX}}

Bitcoin Magazine is another leading and pioneer platform supplying information on the new market. It was founded in Feb 2012, one year earlier than Coindesk. There are fewer news collected from Bitcoin Magazine, only 4,586, but it covers longer period from 28 Feb 2012 to 18 July 2019. Similar to Coindesk, the platform creates channels with different topics like ``Policy and Laws", ``Payment", ``Blockchain" and etc.
But not all news have been classified. We found that 2,976 articles didn't belong to any category. Classification is needed and machine learning methods will be employed. However, our final goal is to apply our methods to multiple data sources, especially the other financial news platforms, such as the NASDAQ news platform in which 887, 018 articles are available since Jan 2013. In order to test the performance of applying our methods to other platforms, we manually marked each article from Bitcoin Magazine with ``policy-related" and ``non-policy-related" to compare with our ML classification results. 582 of 4,586 articles are related to cryptocurrency market regulations.

%
%
%
%
\subsection{CRIX and VCRIX}
The CRIX and VCRIX is chosen to represent the value and the volatility of the entire cryptocurrency market for the later analysis. The CRIX (CRyptocurrency
IndeX), created by \cite{trimborn2018crix}, closely tracks the entire cryptocurrency
market performance. Its construction is robust in the sense it takes into account the dynamics of market structure, thus ensuring the representativity and the tracking performance
of the index. It follows that constituents of CRIX change  over time, depending
on market conditions and the relative dominance of CCs. The CRIX series begins from
July 2014, and is available through thecrix.de. Reallocation of the CRIX happens
on a monthly and quarterly basis. It adopts a liquidity rule when incorporating a certain
cryptocurrency into CRIX, and hence guarantees the trading of CRIX, which is good for ETFs
and traders. CRIX has been widely investigated in the pioneering research on cryptocurrencies,
including \cite{hafner2020testing}, \cite{klein2018bitcoin},  \cite{trimborn2018investing}, and \cite{da2019herding}.

Like VIX or VDAX, which provide a measure for implied volatility, VCRIX, created by \cite{kim2019vcrix}, is a volatility index, able to grasp the risk induced by the cryptocurrency market. This index accurately addresses the market dynamics on the basis of CRIX and thus proved to be a proper basis for option pricing. Similar to CRIX, the data of VCRIX is also be able to be download from thecrix.de.
%
%
%
%
%
%

\section{Methodology}
\label{sec:method}
Based on the rich text corpus, on can now enter the machine learning text mining step to identify policy-related news from others. In this paper, the classification problem is simply binary: policy-related or not. In the literature, SVM is widely applied to solve such kind of binary problem. However, this method didn't performance well with imbalanced cases, whilst our 
target is to classify the regulatory news (a very small subgroup) from all. In our training data, the ratio is $1:7$. Indeed there are multiple ways to solve the unbalanced data problem, such as oversampling or class-weighted SVM by assigning higher misclassification penalties. But the pre-process of oversampling or undersampling will change the distribution of labels and further change the distribution of test data. Our index is constructed based on regulatory news frequency and it is  therefore sensitive to the distribution of classes. 
 
On the other hand, we could assume that when policy-related topics are discussed, similar topics with their key words are used and their distributions are close. Based on that assumption, the Latent Dirichlet Allocation (LDA) method can be employed to analyze the topic distribution and words distribution for the corpus and further identify the policy-related articles according to the similarity calculation. 

\subsection{LDA Topic Modeling}

%
The Latent Dirichlet Allocation (LDA) technique proposed by \cite{blei2003latent}, is an unsupervised machine learning algorithm
that learns the unobserved topics of a corpus
(individual news articles in this paper). This technique is widely applied to establish the thematic structure of text and other discrete data in the linguistic, information retrieval, biologic and even engineering literatures (see \citealp{blei2012probabilistic} for a review of topic modelling and its application to various text collections). 

The LDA technique is based on a generative statistical method to identify the distribution of words that contribute to a topic, while simultaneously constructing documents with different probabilities of topics, meaning that each topic $z$ is annotated with a collections of the most probable words $w$, and each document $d$ is annotated with a collections of the most probable topics $z$. It is an unsupervised algorithm which requires no labeled texts and learns these two latent (unobserved) distributions $p\left(w | z\right)$ and $p\left(z| d\right)$ by acquiring model parameters that maximize the probability of each
word appearing in each document with the number of
topics $K$ as given. 

\begin{figure}[!h]
    \centering
    \includegraphics[width = 12cm]{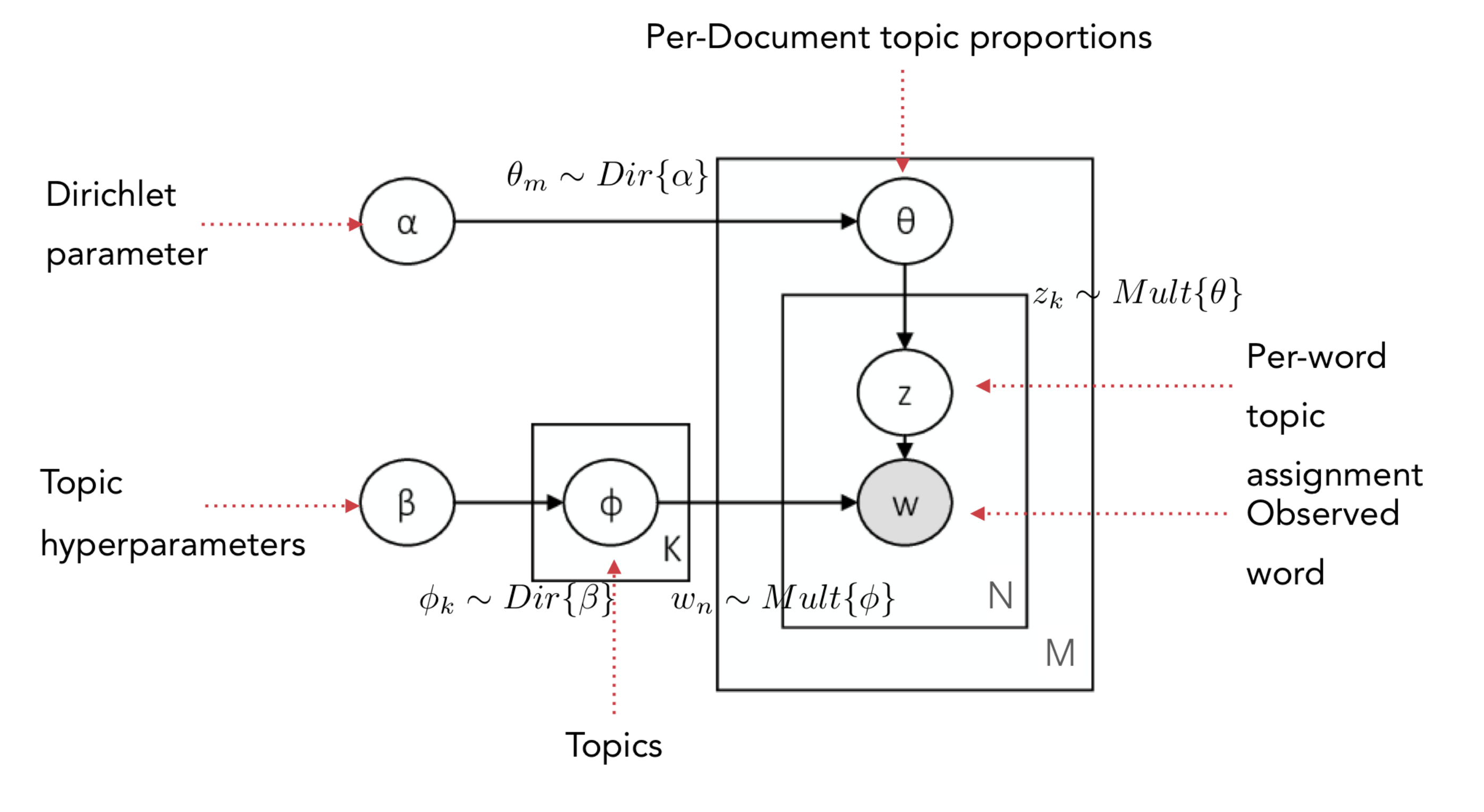}
     \caption[graphic_lda_model]{Graphic LDA Model}
    \label{fig:graphic_lda_model}
\end{figure}

Then, with Bayes theorem, the probability of observed word $w_n$ appearing in a document $d_m$
is given by:

\begin{eqnarray}
p\left(d_{m}, w_{n}\right)&= &
 p\left(d_{m}\right) p\left(w_{n} | d_{m}\right) \\
&= & p\left(d_{m}\right) \sum_{k=1}^{K} p\left(w_{n} | z_{k}\right) p\left(z_{k} | d_{m}\right)
\end{eqnarray}
where $z_k$ is a latent variable indicating the $k$th topic from which
the words were drawn ($Z$ in Figure \ref{fig:graphic_lda_model}), $p\left(w_{n} | z_{k}\right)$ is a distribution for each topic
over the vocabulary ($\phi$ in Figure \ref{fig:graphic_lda_model}), and $p\left(z_{k} | d_{m}\right)$ denotes the topic proportions for the $m$th document (article in this paper) ($\theta$ in Figure \ref{fig:graphic_lda_model}). 
Intuitively, $\phi$ indicates which words weight more to a topic, while $\theta$  states the importance of those topics to a document.


Both $\theta$ and $\phi$ follow the Dirichlet distribution with hyper-parameter $\alpha$ and $\beta$ respectively. With higher $\alpha$, the topic distribution per article turns to be more specific, while similarly, higher $\beta$ leads to a more specific word distribution per topic.  In general, $\alpha$ links to the similarity of documents, meaning that a higher alpha value implies that documents are embodied by more similar weights of each topic. The same holds for $\beta$ but meaning that a higher beta value indicates that topics contents more similar weights of each word.  In the Python package gensim, the symmetric or asymmetric hyper-parameters are learned from data. The generative process of LDA is based on the following joint distribution of the observed variables $w$ 
and the unobserved variables $z$, $\theta$, $\phi$, $\alpha$ and $\beta$,

\begin{equation}
\label{equ:lda}
{\displaystyle p({\boldsymbol {w}},{\boldsymbol {z}},{\boldsymbol {\theta }},{\boldsymbol {\phi }};\alpha ,\beta )=\prod _{k=1}^{K}p(\phi _{k};\beta )\prod _{d=1}^{M}p(\theta _{d};\alpha )\prod _{n=1}^{N}p(z_{d,k}\mid \theta _{d})p(w_{d,n}\mid \phi _{z_{d,k}}),}
\end{equation}

\subsection{Number of Topics for LDA}
The standard LDA model proposed by \cite{blei2003latent} has a significant weakness that it requires pre-determination of the number of topics, meaning that users should set the number of unobserved topics manually before applying the method. The quality of LDA model is heavily depended on the choice of topic number. Therefore, many believe that
choosing the best value for the topic numbers is more art than science (\citealp{azqueta2017developing}).

A common solution is to plug in a set of values and pick the optimal topic number either based on some intrinsic criterion, such as the coherence of the topics, or based on some extrinsic criterion, such as accuracy on a specific task, e.g. paraphrase identification. There also exist other methods to help with choosing the number of topics. For example,  nonparametric Bayesian models, e.g. Hierarchical
Dirichlet process are employed to automatically generate the number of topics
(\citealp{teh2004sharing}). However, it is computationally inefficient to apply such nonparametric models to LDA 
(\citealp{wallach2009rethinking}).

Coherence measures which are based on word co-occurrence are widely applied to quantify the quality of topic models. The poor quality topics with the type of ``chained", ``intruded" and ``random" could be detected using detected with coherence measures (\citealp{mimno2011optimizing}). \cite{newman2010automatic} proposed a coherence measure which is comparable to the human rating of topics. Their coherence measure ($C_{\mathrm{UCI}}$) takes the set of the top J words ($w_1, ..., w_J $) for a given topic and sum a confirmation measure over all word pairs. The function is given as follows:
\begin{equation}
C_{\mathrm{UCI}} =\frac{2}{J \cdot(J-1)} \sum_{i=1}^{J-1} \sum_{j=i+1}^{J} \log \frac{P\left(w_{i}, w_{j}\right)+\epsilon}{P\left(w_{i}\right) \cdot P\left(w_{j}\right)} 
\end{equation}
where the probabilities are estimated on Wikipedia outperform which is used as external reference corpus. \cite{mimno2011optimizing} employ an asymmetrical confirmation measure between top word pairs in the calculation of coherence $C_{\mathrm{UMass}}$:
\begin{equation}
\label{equ:Cumass}
C_{\mathrm{UMass}}=\frac{2}{J \cdot(J-1)} \sum_{i=2}^{J} \sum_{j=1}^{i-1} \log \frac{P\left(w_{i}, w_{j}\right)+\epsilon}{P\left(w_{j}\right)}
\end{equation}
Unlike $C_{\mathrm{UCI}}$, the probabilities in function \ref{equ:Cumass} are estimated based on the original corpus with applied to trained topic models. 
\cite{roder2015exploring} build a coherence framework and report a measure ($C_v$) with the best performance. Different from $C_{\mathrm{UMass}}$ and $C_{\mathrm{UCI}}$, $C_v$ defines the confirmation using normalized point-wise mutual information (NPMI) for the $j-th$ element of the context vector $\overrightarrow{v_{i}}$ of word $w_i$: 
\begin{equation}
v_{ij}=\mathrm{NPMI}\left(w_{i}, w_{j}\right)^{\gamma}=\left(\frac{\log \frac{P\left(w_{i}, w_{j}\right)+\epsilon}{P\left(w_{i}\right) \cdot P\left(w_{j}\right)}}{-\log \left(P\left(w_{i}, w_{j}\right)+\epsilon\right)}\right)^{\gamma}
\end{equation}
where $\gamma$ denotes the weight for NPMI. In this paper, we use coherence value $C_v$ as the criteria to select model. 

\subsection{LDA-Based Similarity Measurement and Classification}

Semantic similarity problems can be classified according to different levels of granularity, specifically ranging from word-to-word to sentence-to-sentence to document-to-document similarities (\citealp{niraula2013experiments}). In this paper, our task is to analyze document-to-document similarity, particularly  as a binary decision problem in which an article is policy related or not. We rely on one probabilistic method, LDA, which regards documents as distribution over topics and topics as distribution over words. So, we assume that policy-related articles have similar topic distributions.  

The Hellinger distance can be applied to compute the distance between two distributions. For document $p$ and document $q$, the distributions of topics are $z_{p}=\left(z_{p, 1}, \dots, z_{p, k}, \dots, z_{p, K}\right)$ and $z_{q}=\left(z_{q, 1}, \dots, z_{q, k}, \dots, z_{q, K}\right)$ respectively. Hellinger Distance $d_H$ between those two news with $K$ topics is given as follows: 

\begin{equation}
d_{H}\left(z_{p}, z_{q}, X\right)=\frac{1}{\sqrt{2}} \sqrt{\sum_{k=1}^{K}(\sqrt{z_{p, k}}-\sqrt{z_{q,k}})^{2}}
\end{equation}

There are reasons that we choose Hellinger distance rather than other distances to calculate news similarity. First, if we denote $\hat{f}(x)$ as a kernel density estimator, the asymptotic distribution of $\sqrt{n h}(\hat{f}(x)-f(x))$ depends on $f(x)$, the true distribution, however, after taking square root, the asymptotic distribution of $ \sqrt{n h} (\sqrt{\hat{f}(x)}-\sqrt{f(x)})$ eliminates its dependency with $f(x)$ (see the proof in Appendix). Second, when applying Hellinger distance in a binary decision criterion, it is not sensitive to the class skew (\citealp{cieslak2008learning}), implying that it performs well with imbalance data. Besides, the results from Hellinger distance is bounded by $[0,1]$ for all values of $ z_{p, k}$ and $ z_{q, k}$. It is easy to read and compare. The highest value, 1, indicates the maximized distance and therefore means that the compared two distributions differ from each other significantly, whereas the value 0 implies the highest similarity and shortest distances. Meanwhile, $d_{H}$ is symmetric, meaning $d_{H}\left(z_{p}, z_{q}\right)$ = $d_{H}\left(z_{1}, z_{p}\right)$.

As the next step, we calculate the average distance between the article $i$ and all policy-related news ${d}_{i}$:

\begin{equation}
\bar{d}_{l,i}=\frac{1}{N_r}\sum_{j=1}^{N_r}d_H(z_{l,i},z_{r,j})
\end{equation}
where $N_r$ is the number of regulatory news, $z_{u,i}$ denotes the topic distribution of article $i$ and $l=\{r, \text{ } non \text{ }or\text{ } u\}$ meaning that the article $i$ is regulatory news ``$r$", non-regulatory news ``$non$" or unclassified news ``$u$". $z_{r,j}$ represents topic distribution of regulatory new $j$ ($j=1,...,N_r$). 

Since we have the assumption that regulatory news have smaller distances between each other than the other news, the average distance $\bar{d}_{r}=\{\bar{d}_{r,1},...,\bar{d}_{r,N_r}\}$ for all policy-related news should be relatively smaller than $\bar{d}_{non}=\{\bar{d}_{n,1},...,\bar{d}_{n, N_{non}}\}$ for all non-policy-related news. Then if $\bar{d}_{u,i}$ of the unclassified article $i$ is small and close to $\bar{d}_{r}$, we mark that news as policy-related. In this paper, we set a threshold $\underline{d}$ equals to $\tau^{th}$ quantile of  $\bar{d}_{r}$ and $\tau=0.95$ in this paper. 

%
%
\subsection{Construction of CRRIX}
As mentioned before, the construction of CRRIX is simply the coverage frequency of policy-related news, as followed: 
\begin{equation}
CRRIX^{s}_t=\frac{N^{s}_{t,reg}}{N^{s}_{t,all}}
\end{equation}

where $s$ is the  periodicity and $s=\{daily, \text{ } weekly \text{ } or\text{ } monthly\}$.
$N_{t,reg}$ and $N_{t,all}$ are the number of regulatory news and all news at time $t$.

\section{Empirical Results }
\label{sec:Results }

\begin{figure}[!h]
    \centering
    \includegraphics[width = 10cm]{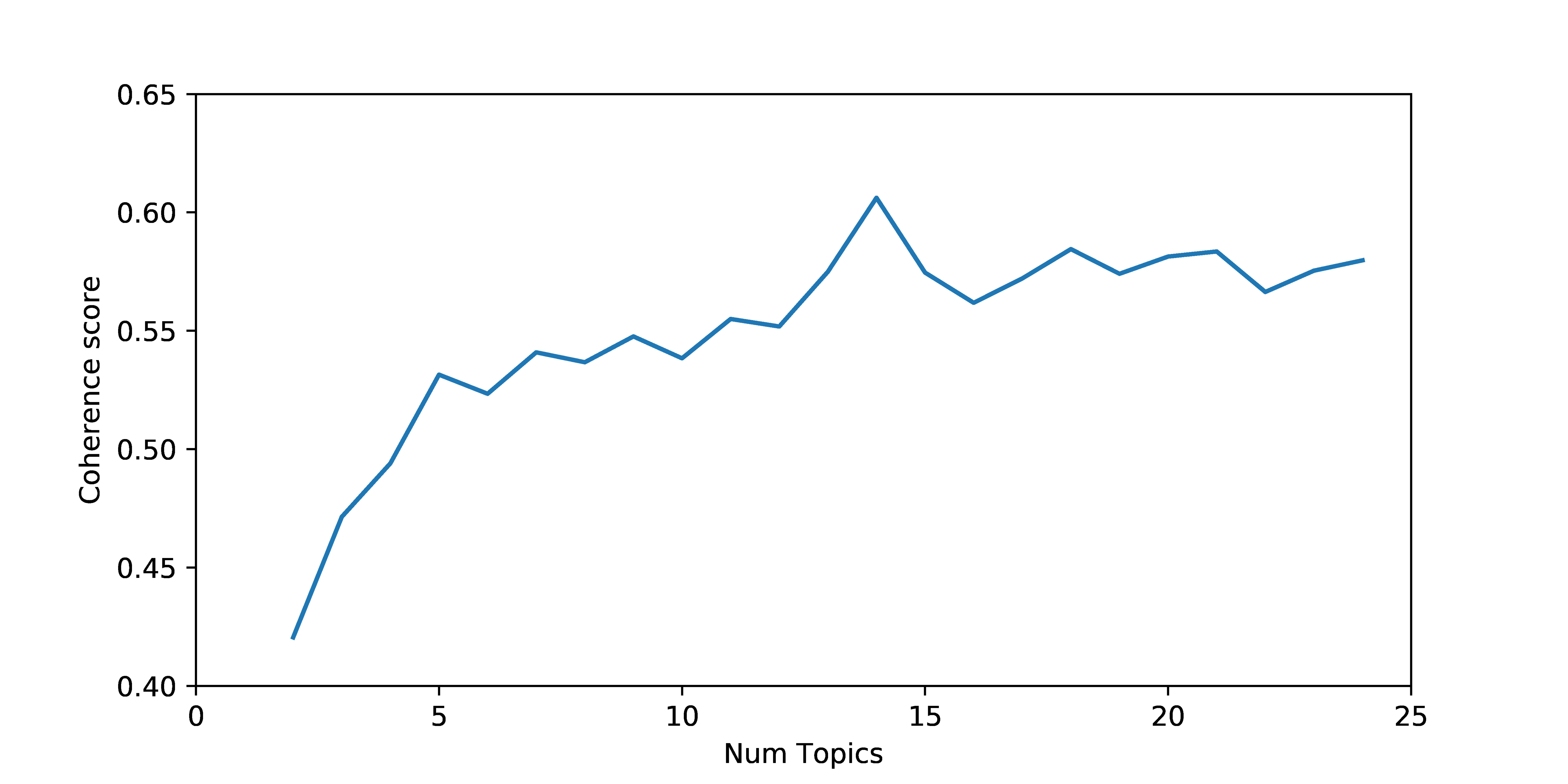}
     \caption[number_of_topics]{ Coherence value for different numbers of topics $K$}
    \label{fig:number_of_topics}
\end{figure}

First we did some pre-processing of the data (words) : stopwords are
eliminated (words that do not informatively or semantically contribute to an article, e.g. ``at", ``or",  ``and"); all words have been
converted to lower case. We calculate the coherence value of LDA models with different topic numbers from $2$ to $25$ given an automatic generated hyperparameter $\alpha$ ($\alpha$=0.01) and $\beta=0.1$. The Figure \ref{fig:number_of_topics} indicates that the best performanced model with optimal number of topics for the corpus in this paper is the model with $K=14$. When $K=14$, the model has the highest coherence value and when the number of topics increases after the optimal choice, the coherence value turns to relatively stable. We also test the robustness with different $\alpha$ and $\beta$ from 0.01 to 0.3. The above mentioned combination performances best but the coherence value doesn't change much for given topic number.  

\FloatBarrier
\href{https://github.com/QuantLet/Regulatory-risk-cryptocurrency}{\qlet{CRRIX}}

\begin{figure}
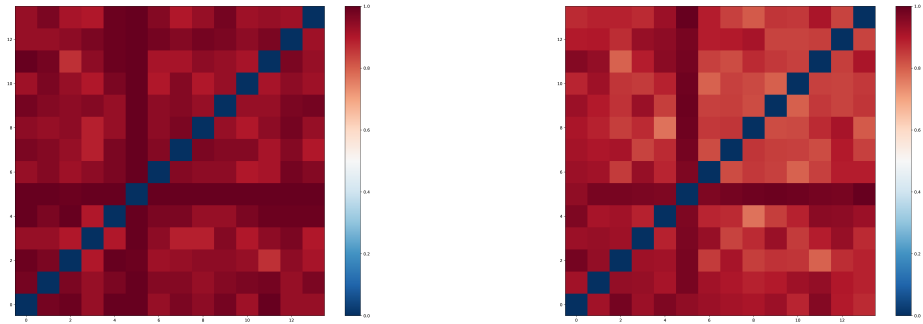

    \centering
    \subfigure{
      \includegraphics[width=7cm]{graphics/topic_distance-eps-converted-to.pdf}
    }
    \subfigure{
      \includegraphics[width=7cm]{graphics/topic_distance_H-eps-converted-to.pdf}
    }
    \caption[distance of topics]{The distances between topics using Jaccard distance (left) and Hellinger distance (right) }
    \label{fig:topic_distance}
\end{figure} 

Another criterion for determining the quality of the LDA model is to concern the diversification of topics. Ideally, we would like to see topics as different as possible. The heating maps in the Figure \ref{fig:topic_distance} exhibit the distances between topics ($K=14$). The red cell represents strongly uncorrelated topics, while the blue cell indicates high correlation. The left diagram was generated using Jaccard distance and for the right one, we apply Hellinger distance to calculate the differences between topics. All elements except those in the diagonal in both diagrams are red or reddish, which means the 14 topics are relatively different from each other and our LDA model performance well in this aspect. However, compared with Hellinger distance, even though Jaccard distance is robust and wildly used in ML methodologies, it is less sensitive than Hellinger distance. Therefore, in the later discussion, we only apply Hellinger's method in the distance calculation.

\begin{table}
\caption{Categories (Coindesk.com) matched by LDA topics}
	\label{table:lda_topics}
	\setstretch{1.5}
	\small
	\begin{adjustbox}{max width = \textwidth}
	\begin{tabular}{ccc}
	
\hline
\hline
\textbf{Coindesk Subcategory}& \textbf{LDA Topic}  & \textbf{Top Keywords} \\\hline\hline
Opinions & Opinions & Bitcoin, say, people, make, go\\ & & get, take, would, could, way\\
\hline
Tech & Technology & System, blockchain, use, transaction, chain\\ & & Technology, security, work, datum, network\\
\hline
Business & Business & Company, say, business, new, service\\ & & base, startup, firm, founder, CEO\\
\hline
Policy \& Regulation & Regulation & Currency, business, virtual, law, state\\ & & Regulation, money, digital, exchange, tax\\
\hline
Market & Investment & Bitcoin, market, currency, price, exchange\\ & & value, investor, Litecoin, trade, investment\\
 & Trading and Exchange & Exchange, BTC, account, customer, trading\\ & & User, deposit, page, trade, fund
\\
\hline
Feature & Mining & mine, power, asic, block, hash\\ & & chip, network, unit, hardware, pool\\
 & Coins & Coin, project, Dogecoin, game, Altcoin\\ & & developer, community, donate, crowdfunder, token
\\
\hline
\hline
\end{tabular}
\end{adjustbox}
\end{table}

In order to further show the performance of the trained LDA model, we try to compare the topics with other sources. 
In the leading Cryptocurrency news platform, Coindesk, the news are labelled as those categories: ``Opinions", ``Tech", ``Business", ``Policy \& Regulations", ``Market" and ``Feature". These categories appear in Table \ref{table:lda_topics} (column 1) together with
their equivalent topic (column 2) which is generated by our LDA model and the list of
representative words for each topic (column 3). From the table we can see that for the major categories in the popular platforms, we could find the corresponding topics in our model. In the case of ``Market", topics go beyond the categories proposed
by the platform.  
We must admit that parts of the category ``Business" overlaps with the category ``Market". Even though we select the topic ``Trading and Exchange" to match the category ``Market", it could also be put under a bigger concept of ``Business". In this sense, the machine learning LDA technique performs better and clearly identifies topics which keep distances with each other.

%

%

We use the trained LDA model to calculate the Hellinger distances. 
We find that the distributions of average distances between each regulatory news and all other regulatory news $\bar{d}_r$ and of average distances between each non-regulatory news and all regulatory news $\bar{d}_{non}$ are significantly different. Policy-related news are similar with  smaller distances, whereas the most non-policy-related news are further away. Then we calculate the average Hellinger distance $\bar{d}_{u,i}$ for each unclassified article $i$. Those, which are smaller than 0.392, the 0.95 quantile of $\bar{d}_r$, will be classified to the policy-related group.  

The classification results of our method based on LDA was compared with those generated by Naive Bayes and SVM, two broadly used supervised ML classification methods. The confusion matrices of classification results and the manually classified data can be found in Table \ref{table:lda_confusion_matrix}. ``True" value is given by human involved annotation, and ``Pred" value is predicted by the here applied ML technique. Number $1$ means that the given news is labelled as policy-related and number those marked by $0$ is non-policy-related.     

\begin{table}[]
\caption{Confusion matrix of classification for three methods (Naive Bayes, Class-weighted SVM and LDA)}
\label{table:lda_confusion_matrix}
\setstretch{1.5}

\begin{tabular}{c| c c| c c| c c |c}
\hline
\hline
\multirow{2}{*}{True\textbackslash Pred} & \multicolumn{2}{c|}{NB} & \multicolumn{2}{c|}{SVM$_{cw}$} & \multicolumn{2}{c|}{LDA} & \multirow{2}{*}{Total} \\

                                         & 1        & 0           & 1         & 0           & 1          & 0          &                        \\
                                         \hline
1                                        & 0        & 582         & 0         & 582         & 361        & 221        & 582                    \\
0                                        & 0        & 4004        & 0         & 4004        & 188        & 3816       & 4004                   \\

Total                                    & 0        & 4586        & 0         & 4586        & 549        & 4037       & 4586 \\
\hline
Accuracy & \multicolumn{2}{c|}{0.873} & \multicolumn{2}{c|}{0.873} & \multicolumn{2}{c|}{0.907} & \\
\hline
\hline           
\end{tabular}
\end{table}

The accuracies of all three methods are relatively high, over 0.87 (seen in Table \ref{table:lda_confusion_matrix}). However, for supervised ML methods Naive Bayes and Class-weighted SVM, they simply label all articles non-policy-related. Even with the high accuracy, those methods can not help with our research question. The purpose of classification in this paper is to find the ratio of policy-related news over all. With the increase of news taken in to the calculation, but 0 policy-related news was identified, the index will go towards destruction.

 Meanwhile, the accuracy of our methods is higher, 0.91. Although from the Table  \ref{table:lda_confusion_matrix} we can read that the type I error for LDA classification are also high, almost $40$ percent, it still could contribute to build the index and in this sense performances much better than NB and SVM.

\begin{figure}[!h]
    \centering
    \includegraphics[width = \textwidth]{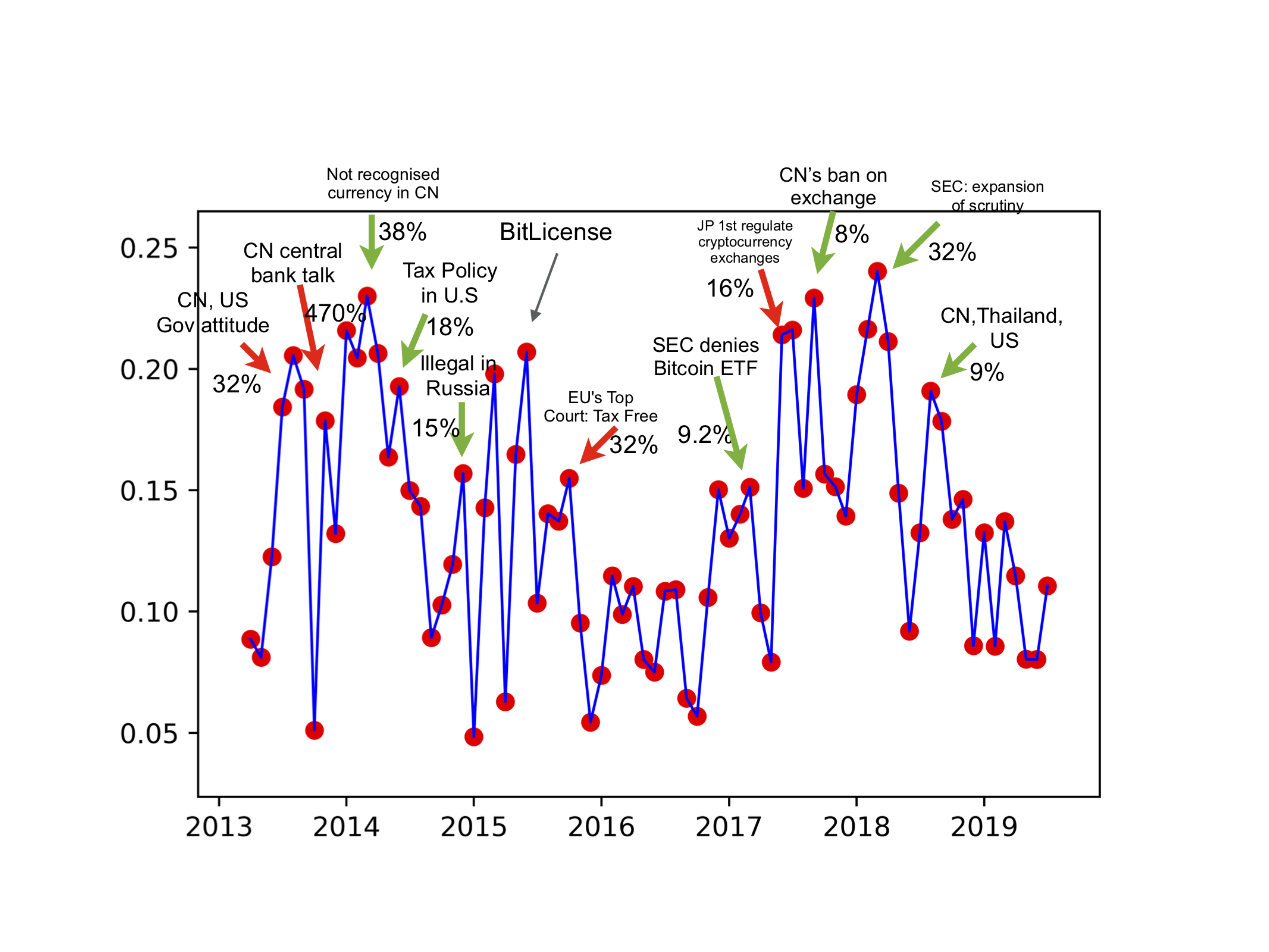}
     \caption[index_results]{CRRIX (Monthly) with news and price highlights}
    \label{fig:index_results}
\end{figure}


Multiple reasons could contribute to the misclassification. One could be that the LDA based criteria are relatively strict. News, which have the distance to all policy-related news as close as that of 95\% the pre-identified regulatory news, would be counted for regulatory news. Those with a bit larger Hellinger distance are all excluded. Another reason could come from the data itself. 
Cryptocurrency market is young and the core policies, which were discussed by the public and announced by the governments, were time-varying. Using all time slot from the year 2013 to 2019 might be biased. A dynamic LDA with rolling window will solve this problem but it requires sufficient data points. In this paper, we didn't consider this method, since the data is limited. 

\FloatBarrier
\begin{figure}[!h]
    \centering
    \includegraphics[width = \textwidth]{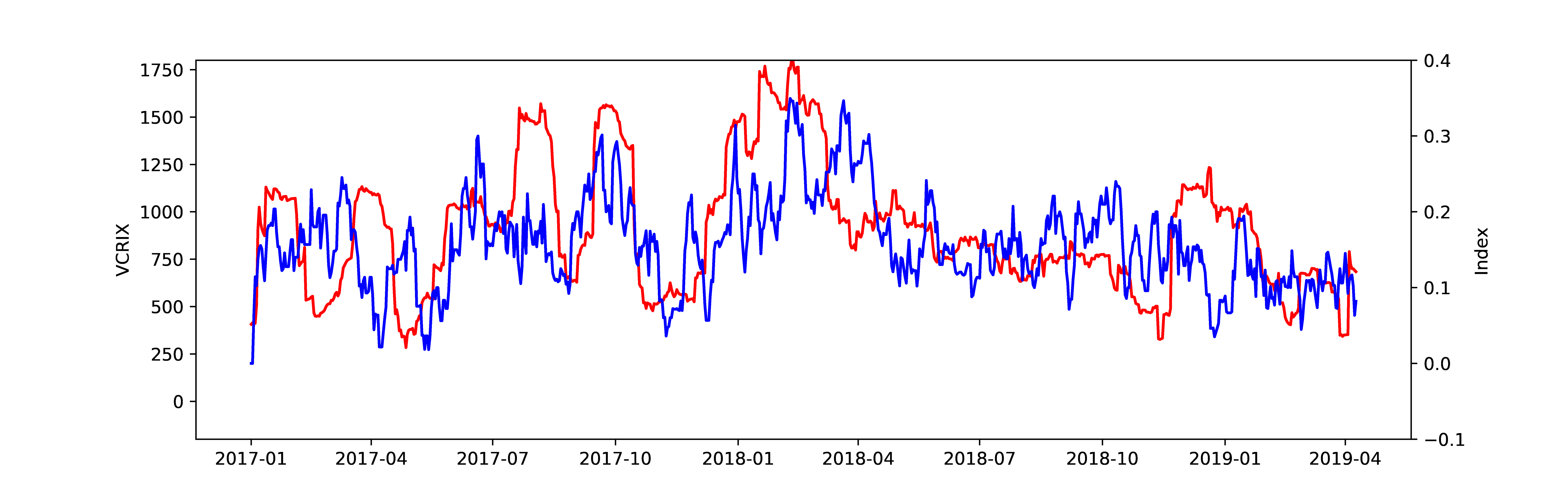}
     \caption[vcrix_index]{VCRIX (in \textcolor{red}{red}) and Regulatory Risk Index of Cryptocurrency Market (in \textcolor{blue}{blue})}
    \label{fig:vcrix_index}
\end{figure}
\href{https://github.com/QuantLet/Regulatory-risk-cryptocurrency}{\qlet{CRRIX}}
\FloatBarrier

After each article was labeled according to its distance with other pre-identified policy-related news, we simply count the number of articles under the class ``regulation" for a give time slot and divide it by the number of all articles for the same time. Since the daily time series is too noisy, especially at the early stage of the market, we only consider weekly time steps. Figure \ref{fig:index_results} indicates that the peaks and jumps of the index are mainly led by big policy changes. The red arrow means positive policy which brought an increase to the price of Bitcoin, whereas the green arrow is vice versa. The number next to arrows is the weekly return rate (positive with red arrow and negative with green arrow). 

Figure \ref{fig:index_results} reveals that the changes of policies are accompanied with drastic price fluctuations which bring high risk to the market. Out index successfully captures those big changing moments.

Figure \ref{fig:vcrix_index} shows that our regulatory risk index is closely related to VCRIX, a volatility index for CCs market. Especially for the period from Sep 2017 to March 2018, the extremely high volatility is driven by the policy uncertainty. The movements for both VCRIX and the regulatory risk index are synchronous.  The correlation between these two indices is 0.44712. Our regulatory risk index could contribute to forecast the market movement. 

We further test the causality between CRRIX and VCRIX. First we do Dicky Fuller test to confirm stationary of both time series. The results reject the non-stationary hypothesis ($p-value$ equals to  $0.000895$ for VCRIX and $0.004617$ for CRRIX). Here we only show the Granger causality test results for lag 1 in the Table \ref{table:gc_result}. The null hypothesis for Granger causality test is that the time series CRRIX, does NOT Granger cause the time series VCRIX. Here, for lag 1, we reject the null hypothesis. This means that the past (lag 1) values of CRRIX (lag 1) have a statistically significant effect on the current value of VCRIX. The results hold for lag 1 to 7. 

\begin{table}
\caption{Granger causality test results for lag $1$}
	\label{table:gc_result}
	\setstretch{1.5}
	\small
	\begin{tabular}{lllll}
	
\hline
\hline

number of lags (no zero) 1\\
\hline
ssr based F test:         & F=23.1736  &p=0.0000  & df\_denom=825 & df\_num=1\\
ssr based chi2 test:   & chi2=23.2579  & p=0.0000  & df=1 &\\
likelihood ratio test: &chi2=22.9372  &p=0.0000   & df=1&\\
parameter F test:         &F=23.1736  &p=0.0000  & df\_denom=825 & df\_num=1\\
\hline
\hline
\end{tabular}
\end{table}

\FloatBarrier
\section{Conclusion}
\label{sec:conclusion}
In this paper, via the machine learning tool LDA, we quantify the risks originating from introducing regulations on the cryptocurrency market and identify their impact on the cryptocurrency investments. 
Indices have been constructed to track the Cryptocurrency markets, however, none of these indices directly address regulatory risks. The indices introduced in \cite{baker2016measuring} focus on economic policy uncertainty in general. Similar to that, we construct a regulatory risk index for  cryptocurrencies that is based on the policy-related news coverage frequency. Unlike the classical annotation approach, which involves a meticulous manual process, we employ the ML-LDA rather costless and efficient method to classify policy-related news. 

We first generally reviewed the development of cryptocurrencies and the trend of regulatory dynamics. Based on that, we discussed the research questions: What exactly is the regulatory risk for Cryptocurrencies?  How to construct an index of regulatory risk for Cryptocurrency market based on news data? What is the impact of regulatory risk to the market?

In order to address the answer to those questions, we first collected news data from the top online cryptocurrency news platform (\citealp{guides2018top}), Coindesk and Bitcoin Magazine, via a dynamic web scraper. 
In addition, the CRIX and VCRIX is chosen to represent the value and the volatility of the entire cryptocurrency market for the later analysis.

To calculate the coverage frequency, we tried to solve the problem of semantic similarity as a binary decision problem in which an article is policy related or not, using Latent Dirichlet Allocation (LDA), which models the underlining topics for a corpus
of documents, where each topic is a mixture over
words and each document is a mixture over topics. 

The topics given by LDA were comparable with the leading Cryptocurrency news platform. For the major categories in the popular platforms, we could find the corresponding topics in our model and the clearly identified topics kept distances with each other. According to our model, the top words for regulation topic are: \textit{currency, business, virtual, law, state, regulation, money, digital, exchange} and \textit{tax}. 

We use the trained LDA model to calculate the Hellinger distances. Those with small average distance will be classified to the group of policy-related news. The results were compared with that of Naive Bayes and Class-weighted SVM methods. Since our data is very imbalanced, the performance of those to supervised ML methods were not helpful. But our LDA based distance classification turned to a high accuracy 0.91 and could contribute to construct the index.

The final results of the regulatory risk index are shown in Figure \ref{fig:index_results} and in Figure \ref{fig:vcrix_index}. Our index successfully captures those big policy changing moments. The movements for both VCRIX and the regulatory risk index are synchronous, and the Granger test proved the causality of CRRIX to the market volatility.

\newpage{}
\clearpage{}

\nocite{*}
\bibliographystyle{aer}
\bibliography{Reg.bib}

\label{appen:sec}
\appendix{}
\newpage{}
\begin{center}
  \begin{huge}
    \textbf{Appendix}
  \end{huge}
\end{center}

\section{Proofs for Hellinger Distance}
\subsection{Asymptotic distribution of $\sqrt{n h}[\hat{f}(x)-f(x)]$ depends on $f(x)$.}
\begin{proof}
Denote $\hat{f}(x)$ is a kernel density estimator, 
\begin{equation}
\sqrt{n h}[\hat{f}(x)-f(x)]=\sqrt{n h}\{\hat{f}(x)-\operatorname{E}[\hat{f}(x)]\}+\sqrt{n h}\{\operatorname{E}[\hat{f}(x)]-f(x)\}
\end{equation}
For the second part, 
\begin{equation}
\begin{aligned} \operatorname{Bias}\left\{\hat{f}_{h}(x)\right\} &=\operatorname{E}\left\{\hat{f}_{h}(x)\right\}-f(x) \\ &=\frac{1}{n} \sum_{i=1}^{n} \operatorname{E}\left\{K_{h}\left(x-X_{i}\right)\right\}-f(x) \\ &=\operatorname{E}\left\{K_{h}(x-X)\right\}-f(x) \\ &=\int \frac{1}{h} K\left(\frac{x-u}{h}\right) f(u) d u-f(x)  
\end{aligned}
\end{equation}
The transformation $s=\frac{u-x}{h}$, i.e. $u=hs+x$, $\left|\frac{d s}{d u}\right|=\frac{1}{h}$. A second-order Taylor expansion of $f(u)$ around $x$ is given by
\begin{equation}
f(x+hs)=f(x)+f(x)^\prime h s+\frac{1}{2}f^{\prime \prime}(x) h^2 s^2 +o\left(h^{2}\right)
\end{equation}
Then,
\begin{equation}
\begin{aligned}
\operatorname{Bias}\left\{\hat{f}_{h}(x)\right\}&= \int \frac{1}{h} K(-s) f(x+hs) h d s-f(x)\\
&= \int  K(s) [f(x)+f(x)^\prime h s+\frac{1}{2}f^{\prime \prime}(x) h^2 s^2 +o\left(h^{2}\right)]  d s-f(x)\\
&= f(x)\int  K(s)ds+f(x)^\prime h \int s K(s)ds+\frac{1}{2}f^{\prime \prime}(x) h^2\int s^2 K(s)ds-f(x)+o\left(h^{2}\right)\\
&=\frac{h^{2}}{2} f^{\prime \prime}(x) \mu_{2}(K)+o\left(h^{2}\right), \quad \text { as } h \rightarrow 0 
\end{aligned}
\end{equation}
where $\int  K(s)ds=1$, $\int s K(s)ds=0$ and  $\int s^2 K(s)ds=\mu_{2}(K)$.

For the first part, 
\begin{equation}
\begin{aligned} \operatorname{Var}\left\{\hat{f}_{h}(x)\right\} &=\operatorname{Var}\left\{\frac{1}{n} \sum_{i=1}^{n} K_{h}\left(x-X_{i}\right)\right\} \\ &=\frac{1}{n^{2}} \sum_{i=1}^{n} \operatorname{Var}\left\{K_{h}\left(x-X_{i}\right)\right\} \\ &=\frac{1}{n} \operatorname{Var}\left\{K_{h}(x-X)\right\} \\ &=\frac{1}{n}\left\{\operatorname{E}\left[K_{h}^{2}(x-X)\right]-\left\{\operatorname{E}\left[K_{h}(x-X)\right]\right\}^{2}\right\}\\
 &=\frac{1}{n} \int \frac{1}{h^{2}} K\left(\frac{x-t}{h}\right)^{2} f(t) d t-\frac{1}{n}\left(\frac{1}{h} \int K\left(\frac{x-t}{h}\right) f(t) d t\right)^{2} \\ &=\frac{1}{n} \int \frac{1}{h^{2}} K\left(\frac{x-t}{h}\right)^{2} f(t) d t-\frac{1}{n}(f(x)+\operatorname{Bias}(\hat{f}(x)))^{2} \end{aligned}
\end{equation}
Substituting $s=\frac{u-x}{h}$, 
\begin{equation}
\operatorname{Var}\left\{\hat{f}_{h}(x)\right\}=\frac{1}{n h} \int K(s)^{2} f(x+h s) d s-\frac{1}{n}\left(f(x)+o\left(h\right)\right)^{2}
\end{equation}
Applying a Taylor approximation yields
\begin{equation}
\begin{aligned}
\operatorname{Var}\left\{\hat{f}_{h}(x)\right\}&=\frac{1}{n h} \int K(z)^{2}\left(f(x)+h s f^{\prime}(x)+\frac{1}{2}f^{\prime \prime}(x) h^2 s^2+o(h^2)\right) d s-\frac{1}{n}\left(f(x)+o\left(h\right)\right)^{2}	\\
&=\frac{1}{n h}\|K\|_{2}^{2} f(x)+o\left(\frac{1}{n h}\right), \quad \text { as } n h \rightarrow \infty	
\end{aligned}
\end{equation}
where $\int  K^2(s)ds=\|K\|_{2}^{2}$.
%
With $\operatorname{Var}(\hat{f}(x)) \rightarrow 0 \text { as } nh \rightarrow \infty$
\begin{equation}
\frac{\hat{f}(x)-\operatorname{E}[\hat{f}(x)]}{\sqrt{\operatorname{Var}(\hat{f}(x))}} \stackrel{d}{\longrightarrow} \mathcal{N}(0,1)
\end{equation}
Substituting the expression for $\operatorname{Var}(\hat{f}(x))$
\begin{equation}
\sqrt{n h}\{\hat{f}(x)-\operatorname{E}[\hat{f}(x)]\} \stackrel{d}{\longrightarrow} \mathcal{N}\left(0, f(x)\|K\|_{2}^{2}\right)	
\end{equation}
If the bandwidth tends to zero faster than the optimal rate, then 
\begin{equation}
\sqrt{n h}\{\operatorname{E}[\hat{f}(x)]-f(x)\} \rightarrow 0
\end{equation}
and the bias term vanishes from the asymptotic distribution,
\begin{equation}
\sqrt{n h}[\hat{f}(x)-f(x)] \stackrel{d}{\longrightarrow} \mathcal{N}\left(0, f(x)\|K\|_{2}^{2}\right)
\end{equation}
\end{proof}

\subsection{Asymptotic distribution of $ \sqrt{n h} [\sqrt{\hat{f}(x)}-\sqrt{f(x)}]$ does not depend on $f(x)$}

\begin{proof}
From transform theorems, we know that if $\text { If } \sqrt{n}(t-\mu) \stackrel{\mathcal{L}}{\longrightarrow} N_{p}(0, \Sigma)$, 
\begin{equation}
\sqrt{n}[f(t)-f(\mu)] \stackrel{\mathcal{L}}{\longrightarrow} N_{q}\left(0, \mathcal{D}^{\top} \Sigma \mathcal{D}\right) \quad \text { for } n \longrightarrow \infty	
\end{equation}
Denote $g(x)=x^{1/2}$, then $\frac{dg}{dx}=\frac{1}{2} x^{-1/2}$.
With $\sqrt{n h}(\hat{f}(x)-f(x)) \stackrel{d}{\longrightarrow} \mathcal{N}\left(0, f(x)\|K\|_{2}^{2}\right)$,
then 
\begin{equation}
\sqrt{n h} [\sqrt{\hat{f}(x)}-\sqrt{f(x)}] \stackrel{d}{\longrightarrow} \mathcal{N}\left(0, \frac{1}{4} \|K\|_{2}^{2}\right)
\end{equation}

The asymptotic distribution of $ \sqrt{n h} [\sqrt{\hat{f}(x)}-\sqrt{f(x)}]$ does not depend on $f(x)$
\end{proof}

\end{spacing}
\end{document}